\documentclass{moriond}
\usepackage{amsmath}

\def\be{\begin{equation}}
\def\ee{\end{equation}}
\def\bea{\begin{eqnarray}}
\def\eea{\end{eqnarray}}

\newcommand{\Photo}{}

\begin{document}
\vspace*{4cm}
\title{Recent Results from NA62 in Kaon and Dump Mode}

\author{J.L. Schubert for the NA62 Collaboration~\footnote{\fontsize{2.5}{4}\selectfont A. Akmete, R. Aliberti, S. Alibocus, F. Ambrosino, R. Ammendola, A. Antonelli, G. Anzivino, R. Arcidiacono, A. Baeva
% ...
}}

\address{Max Planck Institute for Physics\\Boltzmannstr. 8, 85748 Garching, Germany}

\maketitle\abstracts{
    NA62 is a fixed-target kaon experiment at the CERN SPS.
    The recent measurement of the ultra-rare decay $K^+\to\pi^+\nu\bar\nu$, based on data collected between 2016 and 2024 is reported.
    The measurement is compatible with previous measurements and the Standard Model.
    The experiment can also be operated in an alternative beam-dump mode.
    From this mode, a search for new-physics particles with masses in the range of 150 to 2000~MeV, based on data collected in dedicated runs between 2021 and 2024 is reported.
    The search focuses on their decays to $h^\pm\ell^\mp$, where $h^\pm\in(\pi^\pm,\pi^\pm\pi^0,\pi^\pm2\pi^0,K^\pm)$, and $\ell\in(e,\mu)$, which are expected in heavy neutral lepton scenarios.
    No event was observed across all considered signal channels, and upper limits were set on the coupling of such particles to the Standard Model.
}

\section{The NA62 Experiment}

The NA62 experiment is a multi purpose fixed-target experiment at the CERN SPS North Area which covers a broad kaon and beam-dump physics program. 
In the standard kaon operating mode, a $400\,\mathrm{GeV}/c$ proton beam from the SPS impinges on a Be target and produces a secondary unseparated hadron beam (6\,\% $K^+$) of $75\,\mathrm{GeV}$ momentum selected using a set of movable copper-iron collimators (TAX). 

Kaons are identified by a differential Cherenkov counter (KTAG), downstream of which the beam particles are measured by a silicon pixel detector (GTK).
A 117\,m long vacuum vessel contains a 75\,m long fiducial volume (FV) starting 105\,m downstream of the Be target.
The vacuum vessel houses a STRAW spectrometer and is followed by a ring imaging Cherenkov counter (RICH) used for particle identification (PID).
Further PID is performed by a series of calorimeters: the electromagnetic calorimeter (LKr), hadronic calorimeters (MUV1 and MUV2) and a muon detector (MUV3), located behind a 80\,cm-thick iron wall. 
The RICH and the hodoscopes (CHOD) provide time measurements for the charged tracks.
The large angle veto system (LAV) is used to detect final states out-of-acceptance of the calorimeters.
At low angles the calorimeters are complemented by a small angle veto system (SAV). 
A hodoscope (ANTI0) and scintillation counter (CHANTI) located upstream of the FV allow further reduction of background at analysis level. 
The full description of the NA62 beam line and detector is available in reference \cite{NA62:2017rwk}.

\section{Updated branching ratio measurement of $K^+\to\pi^+\nu\bar\nu$ in kaon mode}

The main goal of the NA62 experiment is to measure the ultra rare decay $K^+\to\pi^+\nu\bar\nu$, with a Standard Model (SM) expectation $\mathrm{BR}^\mathrm{SM}_{K\to\pi\nu\bar\nu}=(8.60\pm0.42)\times10^{-11}$ calculated from meson mixing, and $(7.86\pm0.61)\times10^{-11}$ considering a full CKM parameter fit.\cite{Buras:2022wpw,DAmbrosio:2022kvb}
It has been observed above $5\sigma$ significance in NA62 data recorded between 2016 and 2022 with a value $\sim1.7\sigma$ above SM expectation.\cite{NA62:2024pjp}
Building on these results, data recorded in 2023 and 2024 was analysed, corresponding to a normalisation level statistics  increase by a factor $\sim2$ (as measured by the number of normalisation $K^+_{2\pi}$ events). 

By reconstructing a KTAG-tagged Kaon's 4-momentum in the GTK and matching it to a downstream pion, the analysis exploits the missing mass $m_\mathrm{miss}^2= (P_K-p_\pi)^2$, reducing backgrounds from other kaon decay channels by $\mathcal O (10^4)$.
The combined calorimetric and RICH PID guarantee a $\mathcal O(10^7)$ $\mu$ rejection. 
Any event with additional activity in the photon vetos SAV, LAV or LKr is rejected yielding a combined $\mathcal O(10^8)$ $\pi^0$ rejection.
This powerful background rejection, combined with an $\mathcal O(10^{-12})$ single event sensitivity allows NA62 to probe this elusive decay with unprecedented precision.

Several improvements were implemented for the 2023-2024 data-set in terms of hardware, trigger, tracking and PID algorithms, as well as an improved evaluation of the largest source of background in the signal regions (so-called `upstream' background).
For further information see reference~\cite{Chang:2026vvx}.
As a result, the new data-set features a ratio $\sqrt{S+B}/S~\simeq0.25$; a factor 2 improvement with respect to previous analyses.

\begin{figure}
    \centering
    \includegraphics[width=0.45\linewidth]{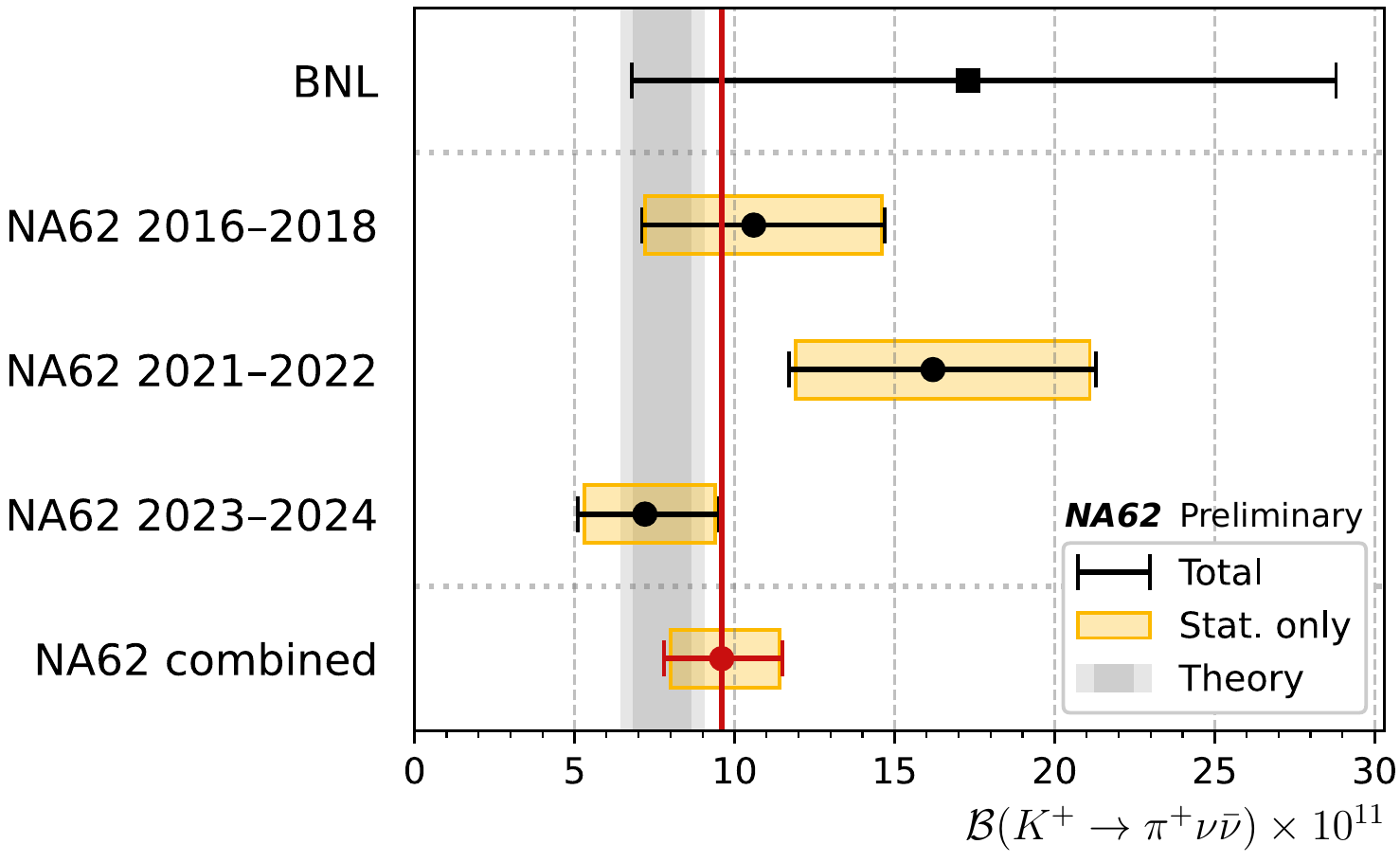}{\hspace{.05\linewidth}}\includegraphics[width=0.365\linewidth]{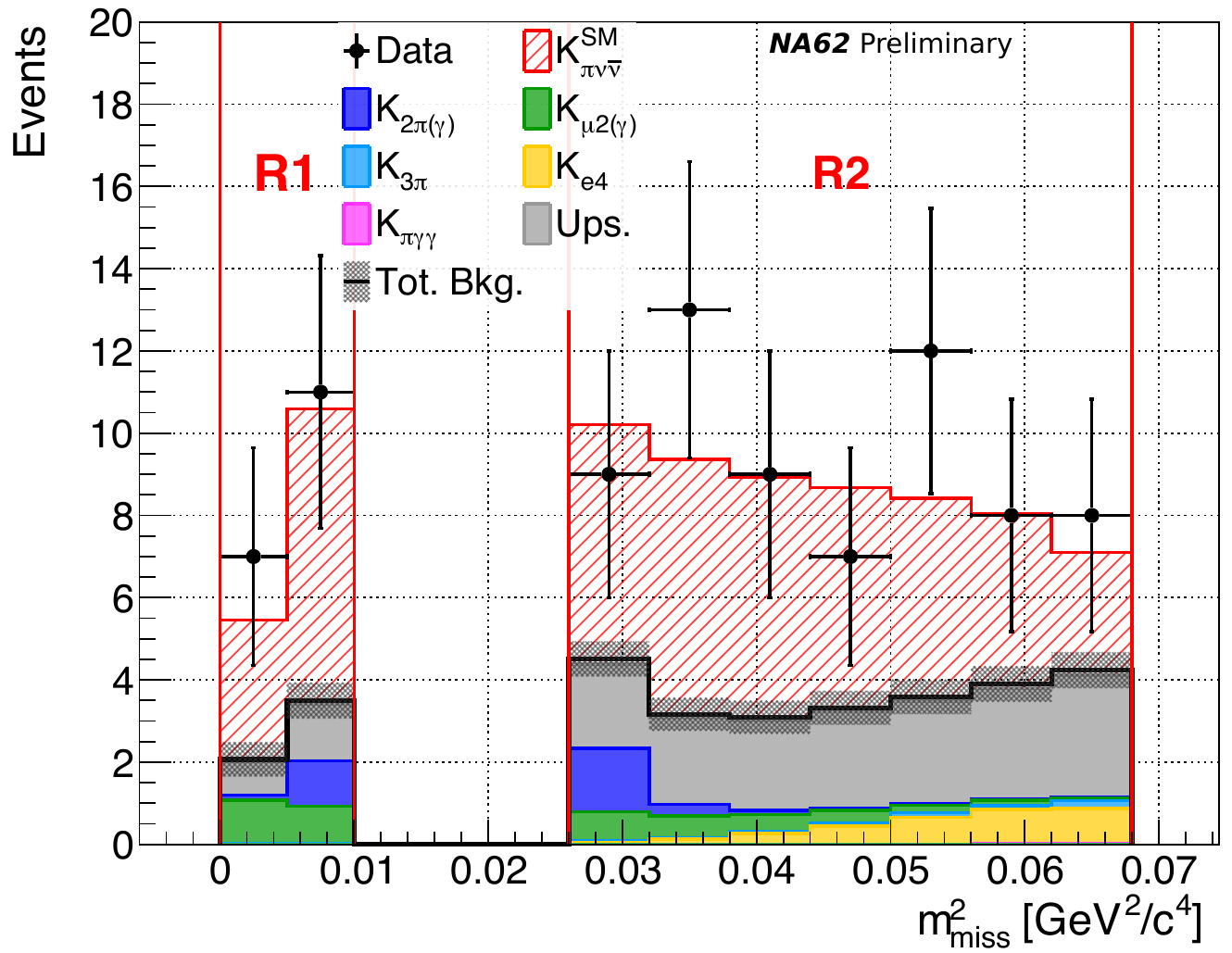}
    \caption{
    Left: Measurements of the $K^+\to\pi^+\nu\bar\nu$ branching fraction and envelope of theory predictions.
    Right: Data-MC comparison of $K^+\to\pi^+\nu\bar\nu$ in Signal Regions R1 and R2 as function of missing mass (assuming $\mathrm{BR}^\mathrm{SM}_{K\to\pi\nu\bar\nu}=8.4\times10^{-11}$).}
    \label{fig:Results_KtoPnn}
\end{figure}

The newly measured branching value $\mathrm{BR}^\mathrm{2023-2024}_{K\to\pi\nu\bar\nu}=(7.2^{+2.3}_{-2.1})\times10^{-11}$ is compatible with previous results reported by NA62, as shown in Fig.~\ref{fig:Results_KtoPnn}(right).
The statistically combined result $\mathrm{BR}^\mathrm{2016-2024}_{K\to\pi\nu\bar\nu}=(9.6^{+1.9}_{-1.8})\times10^{-11}$ is in agreement with SM expectation within $1\sigma$.
This agreement also holds qualitatively at the differential level as shown in Fig.~\ref{fig:Results_KtoPnn}(left) vs. the missing mass $m^2_\mathrm{miss}$.
The result restricts certain Beyond SM scenarios up to scales of 100\,TeV.\cite{Buras:2024ewl}

\section{Search for semileptonic decays in beam-dump mode}

In beam-dump operating mode, the target is lifted, the TAX is closed and used as a dump for the proton beam. 
With a centre of mass energy of $\sqrt{s}=27.3\,\mathrm{GeV}$, it is possible to produce particles with masses above what is kinematically accessible in kaon decays.\cite{NA62:2020mcv}
Between 2021 and 2024, data corresponding to $(6.3\pm1.3) \times 10^{17}$ protons on target was collected during a total of 31 days of operation.

In this mode, NA62 directly probe several long-lived Beyond SM particle scenarios.\cite{NA62:2025yzs}
A search for such long-lived particles decaying to $h^\pm\ell^\mp$, with $h^\pm\in(\pi^\pm,\pi^\pm\pi^0,\pi^\pm2\pi^0,K^\pm)$, and $\ell\in(e,\mu)$ was performed.
The analysis is model-independent, but can be interpreted in terms of a heavy neutral lepton (HNL) $N$ with Yukawa coupling $F_\alpha$ to the SM leptons $L_{\alpha}$ as $\mathcal L\supset-i\sum_{\alpha}F_{\alpha}\bar L_\alpha \sigma_2 H^*N$, where $H$ is the Higgs doublet.
For this interpretation the phenomenological description given by the public tool \textsc{Alpinist} is used.\cite{Schubert:2024hpm}

Signal events in this search are required to feature exactly one charged hadronic and one oppositely charged leptonic track. 
PID is preformed by a classifier developed for this search combining information from STRAW, LKr, MUVs 1--3 and RICH.
This classifier features a $>90\%$ selection efficiency for pions and electrons, and $>99.9\,\%$ for muons.
Additional neutral electromagnetic cluster pairs in the LKr are allowed if their invariant mass is consistent with a $\pi^0$.
Events with any in time activity in the LAV or any in time activity in either ANTI0 or CHANTI geometrically associated with the tracks are vetoed. 

\begin{figure}
    \centering
    \includegraphics[width=0.36\linewidth]{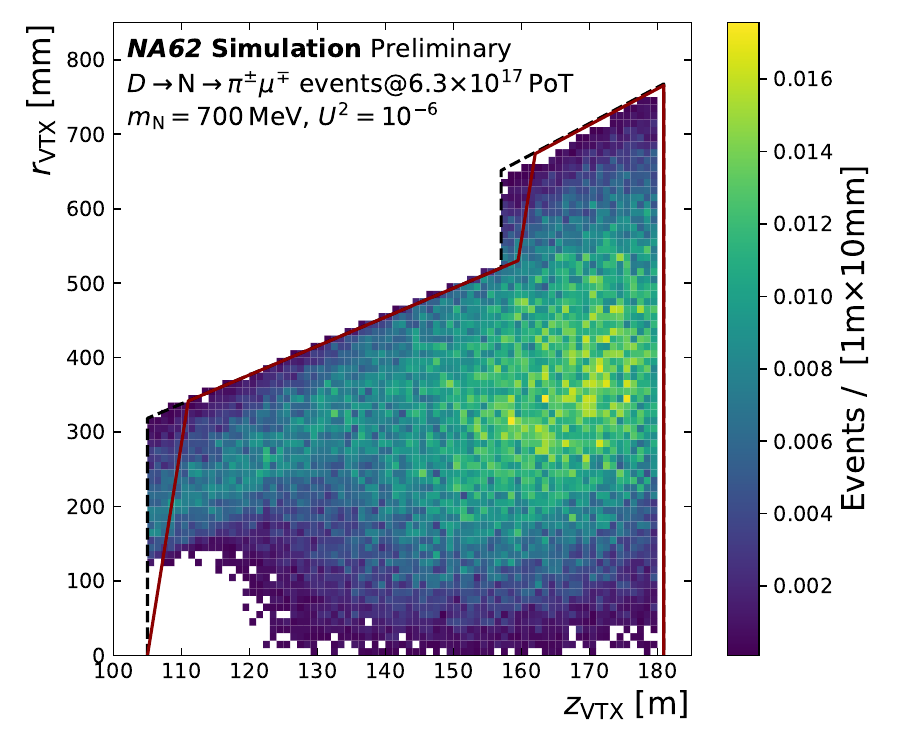}\includegraphics[width=0.36\linewidth]{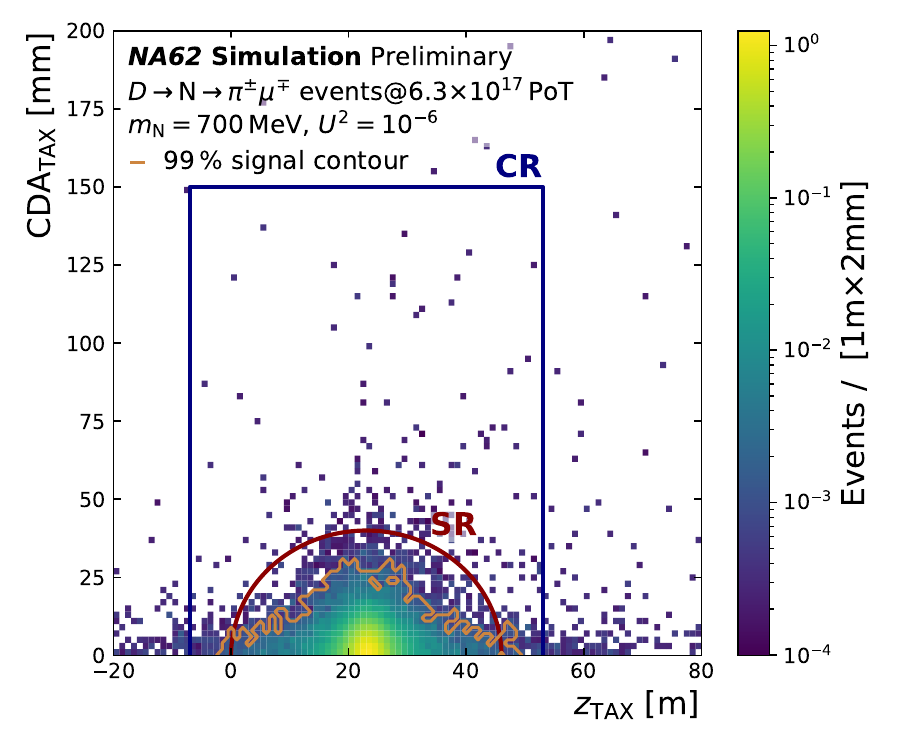}
    \includegraphics[width=0.36\linewidth]{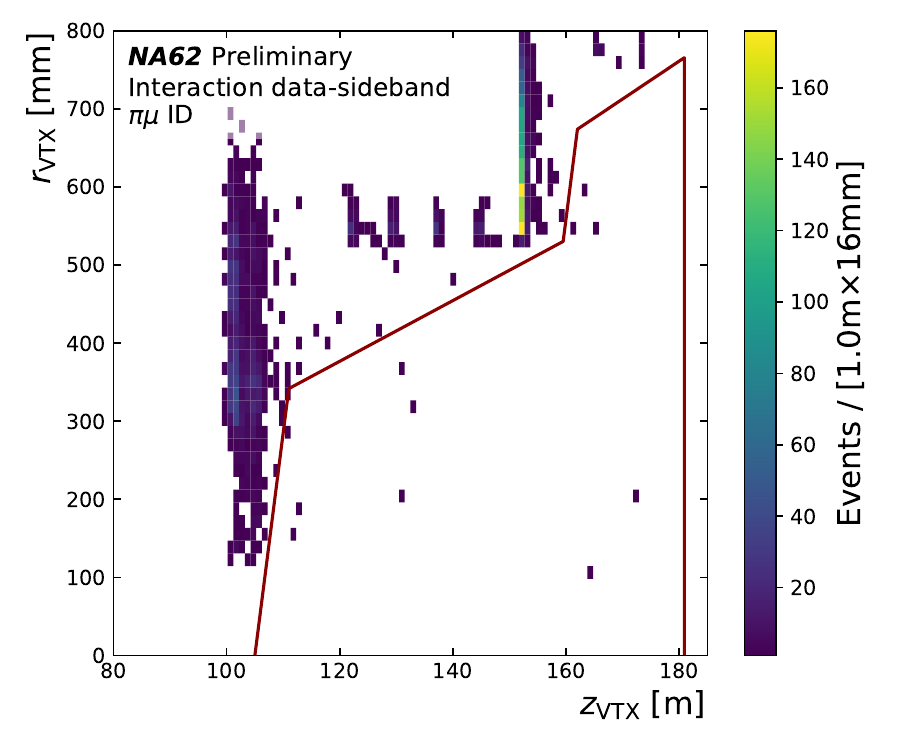}\includegraphics[width=0.36\linewidth]{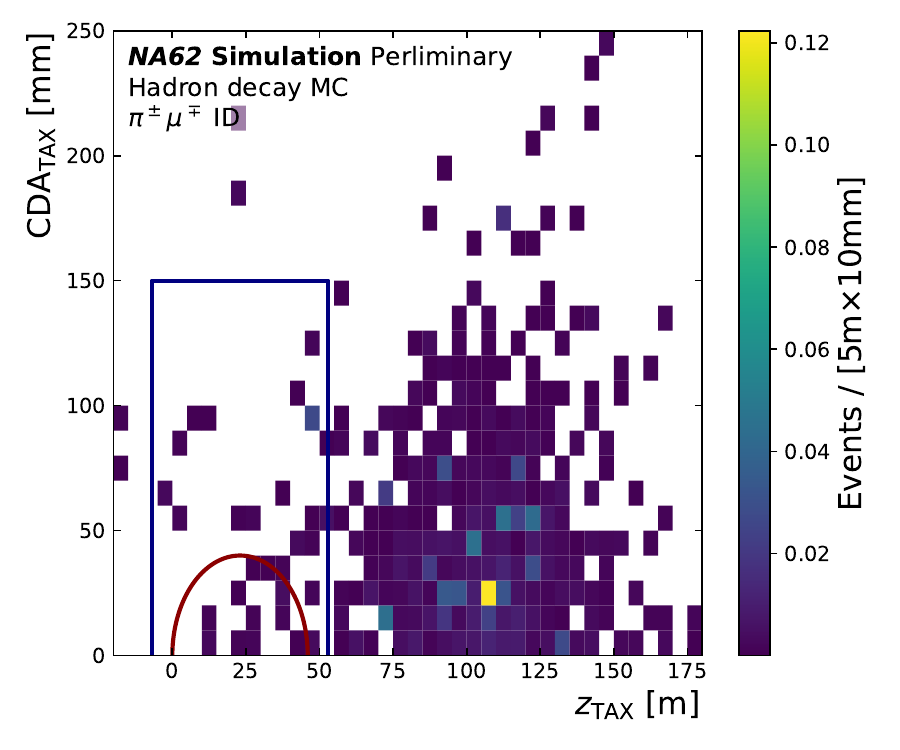}
    \caption{Kinematic separation mechanisms in beam-dump analysis shown for Signal MC with representative parameters (top), and background (bottom). 
    The left plots show the fiducial volume (red) in transversal and longitudinal coordinates of the reconstructed vertex. 
    The right plots show the reconstructed HNL origin with the Signal Region in red.}
    \label{fig:hnl_rejection_mechanisms}
\end{figure}

With the TAX closed, residual charged tracks within the detector are mostly due to single muons.
To suppress background events stemming from interactions of these muons, the tracks must form a high quality vertex inside a fiducial volume subset to the decay vacuum of the NA62 detector. 
Fig.~\ref{fig:hnl_rejection_mechanisms}(left) shows that the signal events are evenly distributed within this FV (top), while background events cluster outside around dense beam line objects such as the final collimator at $z\simeq102\,$m and the LAV stations between $z\simeq120$\,m and $z\simeq175$\,m  (bottom).

A second class of backgrounds is due to decays of hadrons produced upstream and decaying within the fiducial volume. 
Leveraging the closed kinematics of the HNL decay, the likely origin of the HNL can be established, by evaluating its closest point of approach with the the SPS protons' line of flight. 
A Signal Region (SR) is defined in terms of distance ($\mathrm{CDA}_\mathrm{TAX}$) and longitudinal coordinate ($Z_\mathrm{TAX}$) of this point.
This SR remained blinded until the analysis was finalised.
Fig.~\ref{fig:hnl_rejection_mechanisms}(right) shows that most of the signal events lie within the SR (top), while background events cluster around the beginning of the FV at round $z=105\,$m (bottom) with negligible leakage into the SR.

\begin{figure}
    \centering
    \includegraphics[width=0.36\linewidth]{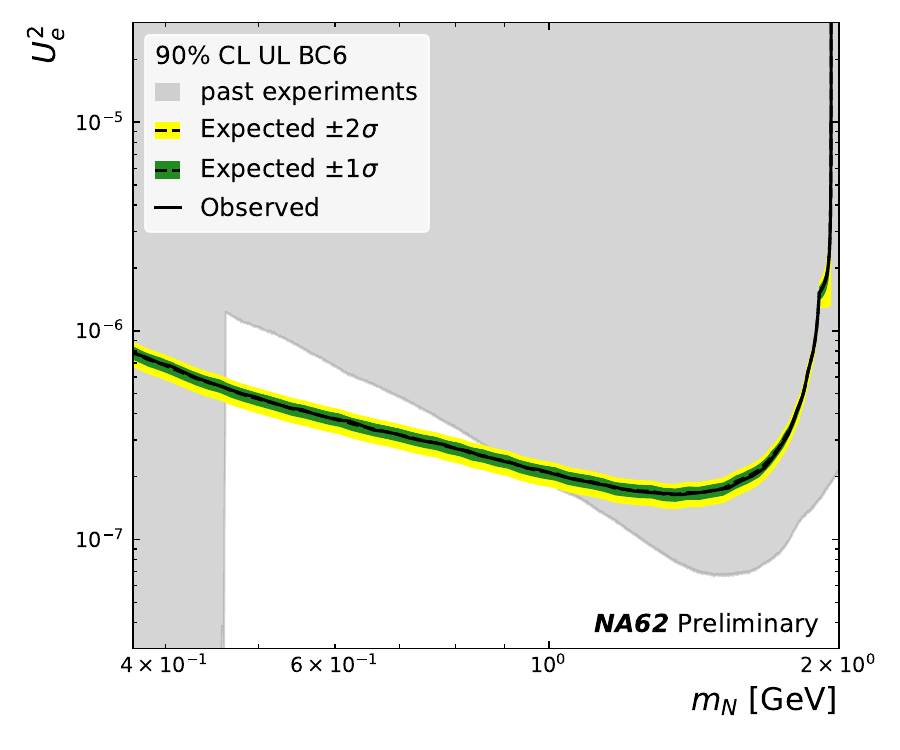}\includegraphics[width=0.36\linewidth]{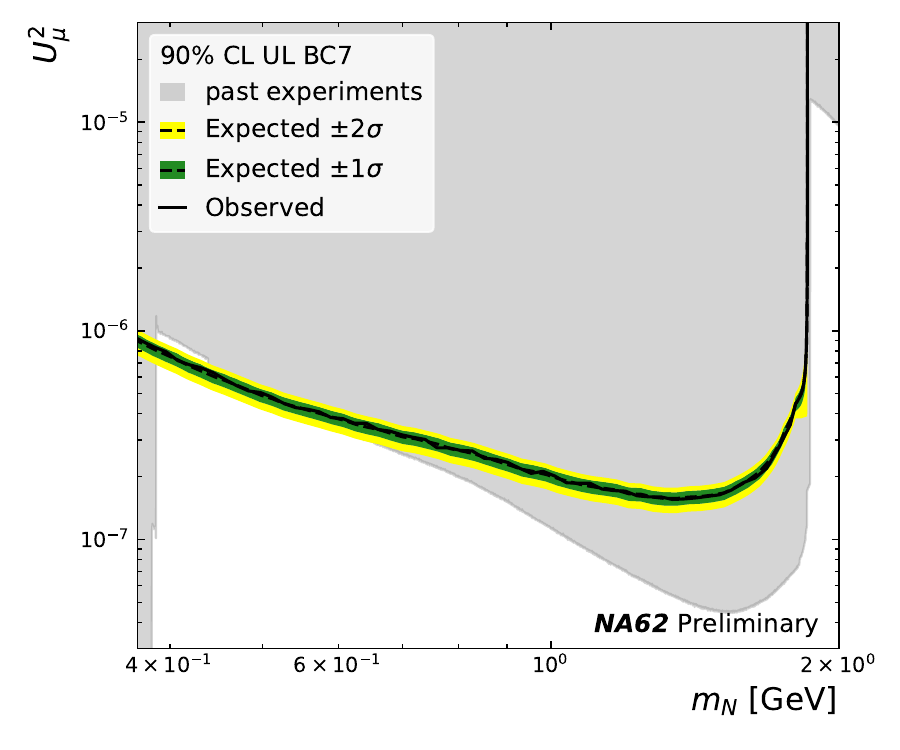}
    \includegraphics[width=0.36\linewidth]{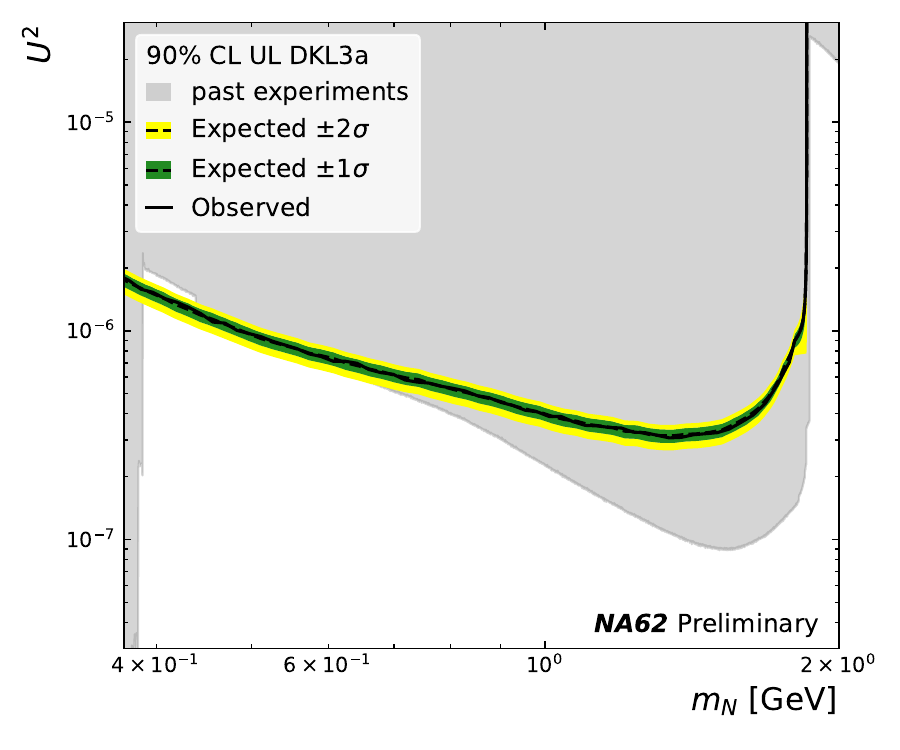}\includegraphics[width=0.36\linewidth]{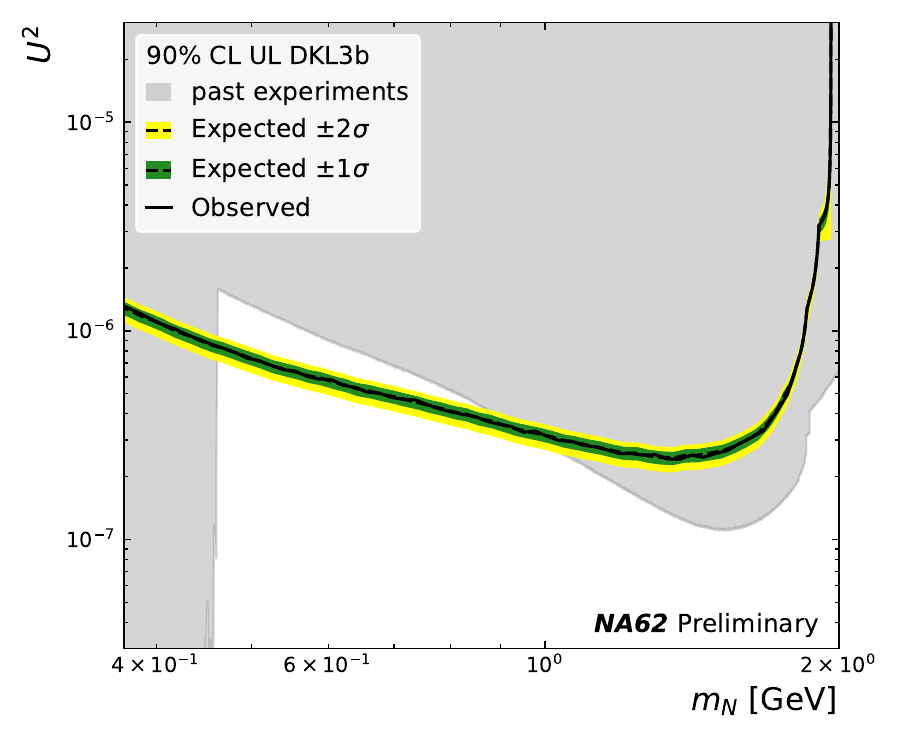}
    \caption{Single Majorana HNL parameter space excluded at 90\,\%\,CL in electrophilic ($U^2_\mu=U^2_\tau=0$, top left), muonphilic ($U^2_e=U^2_\tau=0$, top right), normal hierarchy ($U^2_\mu=U^2_\tau,U^2_e=0$, bottom left), and inverted hierarchy ($U^2_e=U^2_\mu=U^2_\tau$,bottom right) coupling scenarios. The parameter space is reported in terms of HNL mass $m_N$ and coupling suppression $U^2=\sum _\alpha U^2_\alpha$, where $U^2_\alpha=|F_\alpha |^2v/(\sqrt{2}m_N)$.}
    \label{fig:Results_HNL}
\end{figure}

% Accidental combination of unrelated tracks was evaluated to yield a negligible contribution to the background expectation.
No events were observed across any of the signal channels considered. 
The result can be interpreted to exclude parameter space for Heavy Neutral Leptons for standard benchmark cases, as well as neutrino oscillation inspired coupling scenarios.\cite{Agrawal:2021dbo,Drewes:2022akb}
HNLs with coupling suppressions $U^2\sim10^{-6}$ and masses $0.4\,\mathrm{GeV}<m_N<1\,\mathrm{GeV}$ are $90\,\%$ CL excluded as shown in Fig.\ref{fig:Results_HNL}.

% \section*{Acknowledgments}

% I would like to express my sincere gratitude to all members of the NA62 collaboration, who have worked tirelessly to produce these excellent results.

\section*{References}
{\fontsize{11}{5}\selectfont\bibliography{moriondQCD_Schubert}}

%%% manually generated bibliography
%\begin{thebibliography}{99}
%\bibitem{ja}C Jarlskog in {\em CP Violation}, ed. C Jarlskog
%(World Scientific, Singapore, 1988).
%\bibitem{ma}L. Maiani, \Journal{\PLB}{62}{183}{1976}.
%\bibitem{bu}J.D. Bjorken and I. Dunietz, \Journal{\PRD}{36}{2109}{1987}.
%\bibitem{bd}C.D. Buchanan {\it et al}, \Journal{\PRD}{45}{4088}{1992}.
%\end{thebibliography}

\end{document}